\documentclass{revtex4}

\usepackage{graphicx}
\begin{document}
\bibliographystyle{revtex}


\hfill{OUNP-2001-06}

\hfill{December 2001}

\title{Optimising the Linear Collider Luminosity: Feedback on Nanosecond Timescales}



\author{P.N. Burrows}
\email[]{p.burrows@physics.ox.ac.uk}
\affiliation{Oxford University}


\date{\today}

\begin{abstract}
I summarise the R\&D programme on a nanosecond-timescale fast-feedback
system for luminosity optimisation at the linear collider.
\end{abstract}

\maketitle

\noindent
\section{Introduction}

The luminosity achievable in a colliding-beam accelerator may be written:
$$
L\quad=\quad \frac{f \;N\;N_1\;N_2\;H}{4\pi\;\sigma_x\;\sigma_y}
$$
where $f$ is the number of machine cycles per second,
$N$ is the number of bunches per machine cycle, 
$N_1$ and $N_2$ are the number of particles per bunch, and $\sigma_x$, $\sigma_y$
are the transverse overlaps of the colliding bunches. For an $e^+e^-$ collider
$H$ is the beam-beam self-focussing parameter or `pinch enhancement'.
The table shows the values of several of these parameters for the various
designs of a next-generation $e^+e^-$  linear collider currently under consideration.
$\Delta t$ is the time separation between the $N$ bunches in the train.

\begin{table}
\begin{center}
\begin{tabular}{|l|c|c|c|c|c|c|c|}
\hline
Design & Technology & c.m. energy & $f$  & $N$  & $\Delta t$ & $\sigma_x$ & $\sigma_y$ \\
       &            & (GeV)       & (Hz) &      &  (ns)      & (nm)       & (nm)       \\ 
\hline
TESLA  & superconducting & 500-800    &  5   & 2820 & 337        &  553         &  5     \\ 
J/NLC & X-band    & 500-1000        & 120  & 190  & 1.4 &  234         &  4     \\
CLIC   & 2-beam     & 3000        &  75  & 90     &  0.6    &  40          &  0.6         \\
\hline
\end{tabular}
\end{center}
\end{table}

\noindent
The nanometre-level vertical beam overlap is a particularly challenging goal 
for all these designs, most notably CLIC. Any source of beam motion which results
in relative vertical offsets of the two beams at the interaction point (IP) at the
nm level will clearly reduce the luminosity from the nominal value. In all of the
collider designs 
stabilisation below the $1\sigma$ level is required to keep the luminosity loss
below 10\%.

The many kinds of potential beam motion may be characterised in two classes:
(i) slow drifts resulting from eg. thermal excursions or component settling, with
characteristic timescales varying from seconds to months; (ii) jitter on a
timescale comparable with the machine
repetition time. Both kinds of motion were experienced in the decade-long
experience at SLC, and were dealt with by employing slow- and fast-feedback 
systems, respectively. 

We are addressing the design of an intra-bunch-train fast-feedback (FB)
system for the next-generation linear collider (LC).
The system comprises a fast beam position monitor (BPM) to detect the 
relative misalignment of the leading electron and positron bunches at the IP,
a feedback loop, and a fast kicker for kicking the trailing bunches
back into collision. 

The system time-response requirements for J/NLC and CLIC are clearly very different
from those for TESLA. 
We have therefore chosen the more challenging case of J/NLC and CLIC as the primary
focus of our efforts to develop a working prototype hardware system. However, from the 
timing point-of-view, a system which works on the 10ns scale could clearly
be applied to the less demanding TESLA timescale of 300ns.
With current technology  it is clear
that a system for J/NLC and CLIC must be based on fast analogue signal processing, rather
than digital processing~\cite{ingrid}. However, given the rapid advances in signal
processing speeds over the past decade it is quite possible that digital technology
will be fast enough by the time the real LC system is deployed c. 2010; if so, the
FB electronics could look almost identical for any of J/NLC, CLIC or TESLA. 
We aim to monitor this situation and to take advantage of technological developments
as they arise.

\section{Simulation Studies}

During the past 18 months we have created powerful software tools for the
simulation of a nanosecond timescale feedback system for correcting
the relative displacement of nanometre-sized beams.

\begin{itemize}

\item
We have imported and installed the code GUINEAPIG~\cite{schulte}. 
This allows us to
simulate the beam-beam interaction between colliding electron and
positron beams of arbitrary size and bunch charge. When they do not
collide head-on each bunch gives the other a strong transverse EM
kick which can be detected in a downstream BPM. This forms the
physical input to the FB system (Figure 1).

\item
We have installed MATLAB and SIMULINK to create a modular FB system
simulation. The response times of the BPM, feedback loop and  kicker
magnet, as well as cable delays, can all be chosen arbitrarily. The
feedback algorithm can similarly be programmed at will.

\item
We have set up a GEANT model of the NLC interaction region to allow
us to incorporate the FB system components. The location of the BPM
and kicker directly affect the system latency (due to signal 
propagation times). In addition, the material of these components
can contribute to knock-on backgrounds in the detector resulting
from the showering of beam-produced photons and $e^+e^-$.
A corresponding model of the CLIC interaction region is currently
being implemented.

\item
We have set up an equivalent FLUKA model to allow us to calculate
background neutron production.

\end{itemize}

 \begin{figure}
 \includegraphics[width=80mm]{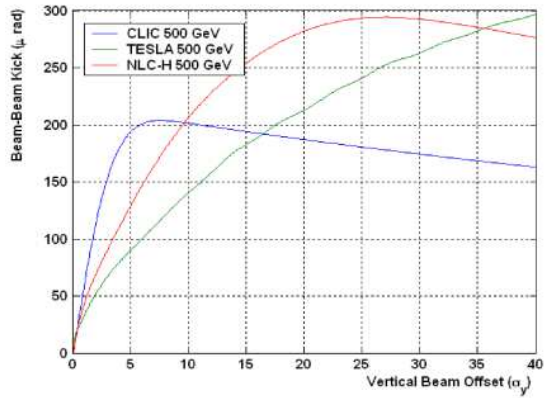}
 \caption{
Simulation of the beam-beam kick that results from beam misalignments
at the IP. This is illustrated for the NLC, TESLA and CLIC machine parameters.
}
 \end{figure}

 \begin{figure}
 \includegraphics[width=80mm]{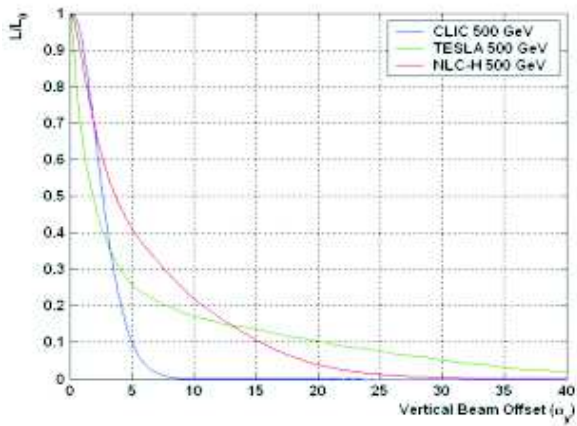}
 \caption{
Fraction of nominal luminosity achieved vs. beam offsets for NLC, TESLA and CLIC,
without feedback.
}
 \end{figure}

\noindent
We have used these tools to simulate the luminosity loss due to beam
misalignment for J/NLC, TESLA and CLIC (Figure 2). For the NLC and TESLA
cases we have set up a feedback model which simulates the performance
of the real hardware components: the BPM, the kicker, the FB logic/loop,
with signal delay times corresponding to the actual locations of the
components within the respective IR. 
In addition, for NLC we have
evaluated the neutron and $e^+e^-$  pair backgrounds in the IR that result
from interactions with the material of the added components; the
equivalent simulations will be done for CLIC.

We have used this software infrastructure to optimise the design of
the FB system for the NLC case in terms of the minimisation of the
response time and the knock-on backgrounds, subject to the constraints
imposed by the locations of the other components in the crowded IR, 
principally the final-focus quadrupole magnets and the vertex detector.
Figure 3~\cite{glen} shows the performance for the optimal hardware layout
in terms of the luminosity loss vs. beam offset and kicker gain. For offsets 
below 10$\sigma$ there is a comfortable `valley' in which the specific
gain choice is not critical. However, for offsets significantly larger
than 10$\sigma$ the choice of gain is delicate; an unfortunate choice could
lead to catastrophic luminosity loss.
We are now in a position to perform similar design optimisation studies
for CLIC and TESLA.

 \begin{figure}
 \includegraphics[width=80mm]{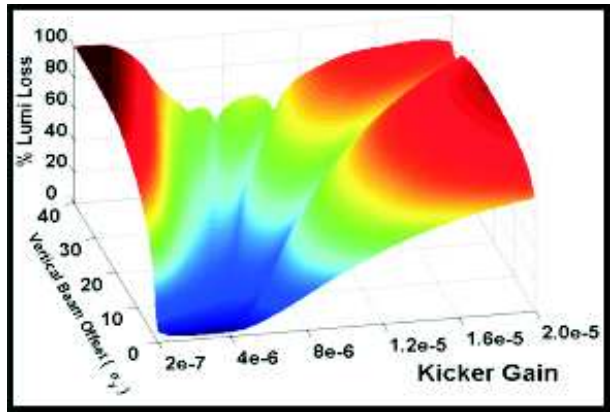}
 \caption{
Simulated luminosity loss for NLC, with feedback. The `valleys'
are a feature of the latency of the system.
}
 \end{figure}

\section{FONT1: First Beam Test at the NLC Test Accelerator}

The importance of FB systems in accelerator stabilisation/control and luminosity
optimisation was learned painfully at SLC. No future linear collider can be
physically aligned for nm-size beam-beam collisions.
The Feedback on Nanosecond Timescales (FONT)
system is key to successful operation of the future linear
collider; without it the luminosity may be 1-2 orders of magnitude
below the nominal design. It is therefore highly desirable to test
experimentally a prototype FB system under conditions as close as
possible to those at the LC~\cite{phil}. 

For J/NLC or CLIC one requires an experiment
with {\it colliding} trains of O(100) bunches with an inter-bunch separation
of O(1 ns); no such facility exists, or is likely to exist, until the collider
turns on. However, {\it non-colliding} bunch trains with similar properties can be
produced. We
have decided to base our first-round test, FONT1, at the NLC Test Accelerator (NLCTA),
located in the Research Yard at SLAC. NLCTA is currently used for the evaluation
of RF components such as Cu structures, but the downstream section is relatively 
`open' beamline.
NLCTA can be operated in a `long-pulse' mode, which provides a train 
of bunches filled at X-band frequency, 11.4 GHz, typically of 120 ns
duration. This
train length is within a factor of 2 of J/NLC and CLIC requirements, and the
total train charge is comparable,
although the bunch spacing, 0.1 ns, is about an order of magnitude shorter; the
prototype experiment is in some sense therefore `harder' than the real case.
We intend to use a dipole magnet to perturb the whole 
train, a fast BPM to monitor the displacement at a given point, an analogue
FB circuit, a fast kicker to attempt to correct the
displacement as fast as possible, and two downstream BPMs to measure and monitor the
correction. 

The key components of the system, and their status, are listed below:

\begin{itemize}

\item
{\bf Dipole magnet:} 
A suitable magnet has been found, a `type 4 linac corrector',
amongst spares at SLAC. It
will produce a dipole offset (i.e. angular deviation) that corresponds to
up to 5 mm position offset at the measurement BPM. The peak current is 6A and it
is planned to use an existing power supply. The dipole can be pre-triggered so
as to be fully `on' by the time the bunch train arrives.

\item
{\bf Fast BPM:}
A prototype X-band `button' BPM has been fabricated. 
The response of the BPM at X-band has been measured using a
network analyser, as well as using the BPM test facility at Lawrence Berkeley Laboratory.
Its resolution has been characterised and appears to be of the order of 10 $\mu$m,
which is more than adequate for our test.

\item
{\bf Fast kicker:}
Three suitable fast kickers have been found amongst
spares at SLAC; only one is needed for a 1-dimensional test.

\item
{\bf Kicker amplifier:}
A preliminary design study for a multi-stage tube amplifier capable of delivering 
up to 5 kW peak power has been made. The design employs Y690 planar triode tubes.

\item
{\bf Diagnostic BPMs:}
Two downstream X-band BPMs (see above) will be used to monitor the
beam position with the FB in off/on modes. 

\end{itemize}

\section{Schedule}

We expect to take data on the FB system performance in spring and summer 2002.
Initially a fixed offset will be applied, and we will attempt to `zero' it as
fast as possible within the bunch train passage time. The fixed offset will
then be varied to simulate different beam offsets in the NLC, 
and we will characterise the performance vs. offset, investigating the effect
of the choice of gain. The ultimate test will be to wire a random-number
generator to the dipole and have the FB system operate `blind'.

Without pre-empting the results of these tests, it is possible to envision
an extended R\&D programme for investigation of a number of issues germane to a real
FB system:

\begin{itemize}

\item
Testing of improved analogue processing algorithms.

\item
Handling of correlated `jitter' within the bunch train, eg. sinusoidal or other
regular transverse position oscillations.

\item
Dynamic gain choice, either from adaptive learning or using a priori
information from upstream systems.

\item
Interplay between position-correction and angle-correction; here we have
addressed only the former.

\item
Correction in both planes: dealing with transverse beam coupling.

\item
Dealing with bunch tails in highly non-Gaussian (i.e. realistic) bunches.

\item
Upgrade to digital technology as signal processing speed improves.

\end{itemize}

Some of these could be carried out at the NLCTA, and/or at other
facilities such as the ATF at KEK, TTF at DESY, or CTF3 which is under construction at CERN.

\end{document}